# A Federated Learning Algorithms Development Paradigm

Miroslav Popovic

University of Novi Sad, Faculty of Technical Sciences, Trg Dositeja Obradovica 6, Novi Sad, Serbia, miroslav.popovic@rt-rk.uns.ac.rs

Marko Popovic

RT-RK Institute for Computer Based Systems, Narodnog fronta 23a, Novi Sad, Serbia, marko.popovic@rt-rk.com

Ivan Kastelan

University of Novi Sad, Faculty of Technical Sciences, Trg Dositeja Obradovica 6, Novi Sad, Serbia, ivan.kastelan@uns.ac.rs

Miodrag Djukic

University of Novi Sad, Faculty of Technical Sciences, Trg Dositeja Obradovica 6, Novi Sad, Serbia, miodrag.djukic@rt-rk.uns.ac.rs

Ilija Basicevic

University of Novi Sad, Faculty of Technical Sciences, Trg Dositeja Obradovica 6, Novi Sad, Serbia, ilija.basicevic@rt-rk.uns.ac.rs

At present many distributed and decentralized frameworks for federated learning algorithms are already available. However, development of such a framework targeting smart Internet of Things in edge systems is still an open challenge. A solution to that challenge named Python Testbed for Federated Learning Algorithms (PTB-FLA) appeared recently. This solution is written in pure Python, it supports both centralized and decentralized algorithms, and its usage was validated and illustrated by three simple algorithm examples. In this paper, we present the federated learning algorithms development paradigm based on PTB-FLA. The paradigm comprises the four phases named by the code they produce: (1) the sequential code, (2) the federated sequential code, (3) the federated sequential code with callbacks, and (4) the PTB-FLA code. The development paradigm is validated and illustrated in the case study on logistic regression, where both centralized and decentralized algorithms are developed.

CCS CONCEPTS • Software and its engineering ~ Software creation and management ~ Software development techniques ~ Software prototyping

## 1 INTRODUCTION

McMahan et al. [1] introduced Federated Learning (FL) as a decentralized model learning approach that leaves the training data distributed on the mobile devices and learns a shared model by aggregating locally computed updates. From the very beginning, Google provided TensorFlow Federated (TFF) [2, 3] as a framework for developing federated learning applications. Many researchers and companies embraced this approach and soon after federated learning became a de facto standard for decentralized model learning in the cloud-edge continuum.

At present many distributed and decentralized frameworks for federated learning algorithms are already available (a short overview is given in section 1.1). However, according to a comparative review and analysis of open-source federated learning frameworks for Internet of Things (IoTs), made by Kholod et al. [4], the application of these frameworks in the IoTs environment is almost impossible. Besides, these frameworks typically have many dependencies, which makes their installation far from trivial, and they are not supported on all the platforms (e.g., TFF and BlueFog are not supported on OS Windows). Therefore, to the best of our knowledge, development of such frameworks is still an open challenge.

Recently, Python Testbed for Federated Learning Algorithms (PTB-FLA) [5] was offered as a a framework for developing federated learning algorithms (FLAs) i.e., as a runtime environment for FLAs under development, on a single computer (and on edge systems in the future). PTB-FLA was written in pure Python to keep application footprint small (to fit to IoTs) and to keep its installation simple (with no external dependencies).

PTB-FLA programming model is a restricted programming model, which imposes the following two restrictions: (1) using the Single Program Multiple Data (SPMD) pattern, and (2) specifying code for server and client roles in form of callback functions. Enforced by these restrictions, a developer writes a single application program, which is instantiated and launched by the PTB-FLA launcher as a set of independent processes whose behaviour depends on the process id. During processes execution, the callback functions are called by the generic federated learning algorithms hidden inside PTB-FLA. PTB-FLA supports both centralized and decentralized federated learning algorithms, and its usage was validated and illustrated in [5] by three simple algorithm examples (a more detailed overview of PTB-FLA is given in section 2).

On the other hand, the main limitations of the paper [5] are that it falls short on providing: (1) a more systematic approach to development of FLAs and (2) an example of a commonly used ML algorithm. This paper is a follow up paper on [5], which is motivated by the desire to overcome the previously mentioned limitations.

In this paper, we present the federated learning algorithms development paradigm based on PTB-FLA. The main goals of PTB-FLA development paradigm are: (1) to aid the development of FLAs that are correct by construction and (2) to make the development of FLAs easier. Therefore, PTB-FLA development paradigm is devised as a series of program code transformation phases where each transformation phase consumes its input code and produces the semantically equivalent code that is closer to the target PTB-FLA code. Here semantically equivalent means that the codes produce the same output data. By convention, the phases are named according to their output codes. Altogether, the PTB-FLA development paradigm comprises the four phases: (1) the sequential code, (2) the federated sequential code, (3) the federated sequential code with callbacks, and (4) the PTB-FLA code.

The PTB-FLA development paradigm is validated and illustrated in the case study on logistic regression, which is one of the simplest and at the same time one of the most used Machine Learning (ML) algorithms. In this case study, we train the logistic regression model that can make predictions on clients' orders based on their profiles. In the case study, we apply the PTB-FLA development paradigm to construct both centralized and decentralized logistic regression FLAs.

In summary, the main contributions of this paper are: (1) the PTB-FLA development paradigm, (2) the case study on logistic regression, and (3) the developed centralized and decentralized logistic regression FLAs.

The rest of the paper is organized as follows. The section 1.1 presents related work, the section 2 presents a short PTB-FLA overview, the section 3 presents the PTB-FLA development paradigm, the section 4 presents the case study on logistic regression, and the section 5 presents the paper conclusions.

## 1.1 Related Work

This section presents a brief overview of the most closely related research that was conducted before this paper.

Back in 2017, *federated learning* was introduced by McMahan et al. [1] as a decentralized approach to model learning that leaves the training data distributed on the mobile devices and learns a shared model by aggregating locally computed updates. They presented FedAvg, a practical method for the federated learning of deep networks based on iterative model averaging (see algorithm FederatedAveraging in [1]). The main advantages of federated learning are: (1) it preserves local data privacy, (2) it is robust to the unbalanced and non-independent and identically distributed (non-IID) data distributions, and (3) it reduces required communication rounds by 10–100x as compared to synchronized stochastic gradient descent (FedSgd).

Immediately after the McMahan's seminal paper [1], federated learning got its traction. Widespread research in both industry and academia resulted in many researchers' papers, and in this limited space we mention just few of them. Not long after [1], Bonawitz et al. [6] introduced an efficient secure aggregation protocol for federated learning, and Konecny et al. [7] presented algorithms for further decreasing communication costs. More recent papers are more focused on data privacy [8, 9].

TensorFlow Federated (TFF) [2], [3] is Google's framework supporting the approach introduced in [1], which provides a rich API and many examples that work well in Colab notebooks. However, TFF is a framework for applications in the cloud-edge continuum, with a heavyweight server executing in the cloud, and therefore not deployable to edge only. Besides, TFF is not supported on OS Windows, which is used by many researchers, and TFF has numerous dependencies that make its installation far from trivial.

BlueFog [10], [11] is another federated learning framework with the same limitations as TFF: (1) BlueFog is cloud dependant because BlueFog authors consider deep training within high-performance data-centre clusters, see note on page 5 in [11] and (2) BuleFog has many dependencies on external software packages and is not supported on OS Windows.

Recently, Kholod et al. [4] made a comparative review and analysis of open-source federated learning frameworks for IoT, including TensorFlow Federated (TFF) from Google Inc [6], Federated AI Technology Enabler (FATE) from Webank's AI department [12], Paddle Federated Learning (PFL) from Baidu [13], PySyft from the open community OpenMined [14], and Federated Learning and Differential Privacy (FL&DP) framework from Sherpa.AI [15]. They concluded that based on the results of their analysis, currently the application of these frameworks on the Internet of Things (IoTs) environment is almost impossible. In summary, at present, developing a federated learning framework targeting smart IoTs in edge systems is still an open challenge.

TensorFlow Lite [16] is a lightweight solution for mobile and embedded devices, which enables both on-device training and low-latency inference of on-device machine learning models with a small binary size and fast performance supporting hardware acceleration [17]. So, TensorFlow Lite is not a federated learning framework, but it is an orthogonal AI framework for mobile devices, which might be combined with a federated learning framework such as PTB-FLA and this is one of the possible directions of our future work.

PyTorch Mobile [18] (formerly PyTorch Lite) is another AI framework very similar to TensorFlow Lite, which provides an end-to-end workflow, from training a model on a powerful server to deploying it on a mobile device, while staying entirely within the PyTorch [19] ecosystem. This simplifies the research to production and paves the way for federated learning techniques. Luo et al. [20] made a comprehensive comparison and benchmarking of AI models and AI frameworks (PyTorch Mobile, Caffe2 which is now part of PyTorch Mobile, and TensorFlow Lite) on mobile devices, and concluded that there is no one-size-fits-all solution for AI frameworks on mobile devices (see remark 2 on page 8 in [20]), because TensorFlow Lite performs better for some AI models or devices whereas PyTorch Mobile performs better for other AI models or devices.

Finally, we would like to clarify what PTB-FLA is not and why it is called a "testbed". PTB-FLA is neither a complete system such as CoLearn [21] and FedIoT [22] nor a system testbed such as the one that was used for testing the system based on PySyft in [23]. By contrast, PTB-FLA is just a FL framework, which is seen by ML&AI developers in our project as an "algorithmic" testbed where they can plugin and test their FLAs.

The PTB-FLA programming model is based on the SPMD pattern [24] like other well-known programming models: MapReduce, MPI, OpenMP, and OpenCL. For those who know the MapReduce programming model it doesn't take long to realize that the PTB-FLA and MapReduce are rather similar – the client callback function in PTB-FLA is like the map callback function in MapReduce, whereas the server callback function in PTB-FLA is like the reduce callback function in MapReduce.

Next, we briefly discuss our design choice to run the application instances as processes. Those who are involved in server applications for servicing many user requests, frequently use actors rather than processes, because they are lightweight such that many actors can run on a single thread. So, why don't we use actors? The first and the main reason is that unlike the server applications where many actors run in the same server, the instances of PTB-FLA applications would normally run on different smart IoTs, each instance on its own device. As already mentioned, we plan to extend PTB-FLA to support this distributed execution model in our nearest future work.

The second reason for not considering actors is because they are not supported in pure Python. The Pythonic alternative to actors is Python asyncio abstractions, and we might consider introducing them in our future work when needed. While discussing system overheads (performance), it is worth mentioning that PTB-FLA is also targeting embedded platforms without the OS, where a single application runs on a bare machine or some runtime/interpreter, such as for example MicroPython [25] boards where MicroPython (Python 3, written in C and optimized to run on a microcontroller) is the replacement for the OS.

Another work highly related to system performance is Codon [26], a compiler for high-performance Pythonic applications and Domain Specific Languages (DSLs). In broader sense, Codon is a full language (and compiler) that borrows Python 3's syntax and semantics (with some limitations) and compiles to native machine code with low runtime overhead, allowing it to rival C/C++ in performance. Besides providing great performance, Codon is very interesting for the TaRDIS project, because it uses a novel bidirectional Intermediate Representation (IR) and compiles it to LLVM IR that is in Static Single-Assignment (SSA) form (SSA-form). This makes it possible to construct a tool that translates SSA-form into a behavioral model, which then may be used to automatically check desired system properties (safety, liveness, security, privacy, etc.).

Logistic regression is an important ML technique for analyzing and predicting data with categorical attributes, and in our case study (see section 4) we took the simple implementation of the logistic regression at [27] as the source for our referent sequential code. In our future work, we plan to consider more advanced models, such as the generalized linear model, see [28].

## 2 PTB-FLA OVERVIEW

This section presents the short PTB-FLA overview (for more details see [5]). We use the term PTB-FLA system to refer to a system based on PTB-FLA. The next three subsections present the PTB-FLA system architecture, PTB-FLA API, and PTB-FLA system operation, respectively.

### 2.1 PTB-FLA System Architecture

The PTB-FLA system architecture comprises the application launcher process $s$, the distributed application $A$, which is a set of application program instances $a_i$, and the distributed testbed $T$, which is a set of testbed instances $t_i$, where $i = 1, 2, …, n$, and $n$ is the number of instances in both $A$ and $T$.

The block diagram of PTB-FLA system architecture is shown in Figure 1. The launcher process $s$ (see top of Figure 1) instantiates $n$ application program instances $a_i$, $i = 1, 2, …, n$, and launches them as $n$ independent processes (see the outer orbit in Figure 1). Each application program instance $a_i$ in turn creates its testbed instance $t_i$ (see the inner orbit in Figure 1). The testbed instances complete the startup procedure by exchanging hello messages.

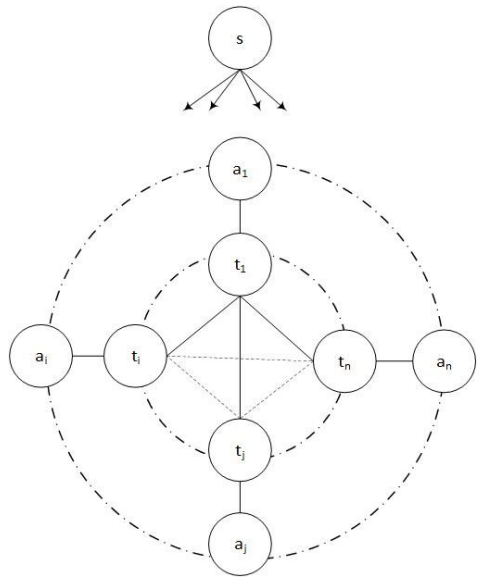

Figure 1: Block diagram of the PTB-FLA system architecture.

The distributed federated learning algorithm is executed as follows. Each instance $a_i$ prepares its input data based on its command line arguments (including its identification $i$, the number of instances $n$, etc.) and then calls the generic API function fl_centralized (in case of a centralized FLA) or the API function fl_decentralized (in case of a decentralized FLA) on its testbed instance $t_i$. Further on, all testbed instances take part in the generic algorithm by exchanging messages with other testbed instances and by calling the associated callback functions at the right points of the generic algorithm. In case of a centralized algorithm, the graph of testbed instances takes the form of a star (see solid edges in the centre of Figure 1), whereas in the case of a decentralized algorithm it takes the form of clique (see solid and dashed edges in the centre of Figure 1).

The PTB-FLA system architecture comprises three layers: the distributed application layer on top (including the application modules and the console script launch), the PTB-FLA layer (including the class PtbFla in the module ptbfla and the modules mpapi and launcher), and the Python layer (including classes Process, Queue, Listener & Client from the package multiprocessing and Popen from the package subprocess), see Figure 2.

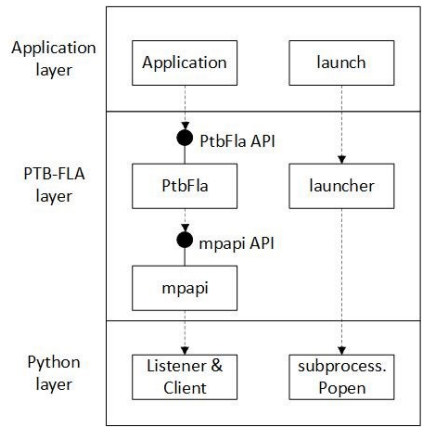

Figure 2: UML class diagram of the PTB-FLA system architecture.

The application module uses the PtbFla API (comprising PtbFla functions) to create or destroy a testbed instance (by calling the constructor or the destructor) and to conduct its role in the distributed algorithm execution (by calling the API function fl_centralized or the API function fl_decentralized). The API functions fl_centralized and fl_decentralized, within an instance $t_i$, use the module mpapi (mpapi is the abbreviation of the term *message passing API*) to communicate with other instances by using the classes Listener and Client in the Python layer (for more details on the PTB-FLA system architecture see [5]).

By design, the mpapi API is strictly an internal API providing services to PtbFla only, and by convention the distributed algorithms' developers should always only use the PtbFla API in their application program modules.

## 2.2 PtbFla API

The PtbFla API offers the following four functions (the function return value is specified as its type):

1. None  PtbFla(*noNodes*, *nodeId*, *flSrvId*=0)
2. *ret* fl_centralized(*sfun*, *cfun*, *ldata*, *pdata*, *noIterations*=1)
3. *ret* fl_decentralized(*sfun*, *cfun*, *ldata*, *pdata*, *noIterations* =1)
4. None PtbFla()

The first and the last are the constructor and the destructor, respectively, whereas the second and the third are the generic centralized FLA and the generic decentralized FLA.

The arguments are as follows: *noNodes* is the number of nodes, *nodeId* is the node identification, *flSrvId* is the server id (default is 0; used by the function fl_centralized), *sfun* is the server callback function, *cfun* is the client callback function, *ldata* is the initial local data, *pdata* is the private data, and *noIterations* is the number of iterations that is by default equal to 1 (for the so called one-shot algorithms). The return value *ret* is the node final local data. Data (*ldata* and *pdata*) is application specific. Typically, *ldata* is a machine learning model, whereas *pdata* is a training data that is used to train the model.

## 2.3 PTB-FLA Operation

As already mentioned in the previous section, the PtbFla API functions fl_centralized and fl_decentralized are the generic centralized FLA and the generic decentralized FLA, respectively. This section briefly explains the system operation governed by these functions.

The common part for both centralized and decentralized FLAs is the interaction between the server $a_i$ and the client $a_j$, which from the $a_i$ point of view proceeds through three phases, see Figure 3, as follows. In the phase 1, the server $a_i$ sends its local data to the client $a_j$, in the phase 2, the server $a_i$ receives the client $a_j$ update (i.e., updated data), and in the phase 3, the server $a_i$ calls its callback function to aggregate all the updates and update its local data. On the other hand, once the client $a_j$ receives the local data from the server $a_i$, it calls its callback function to get its update and then sends the update to the server $a_i$.

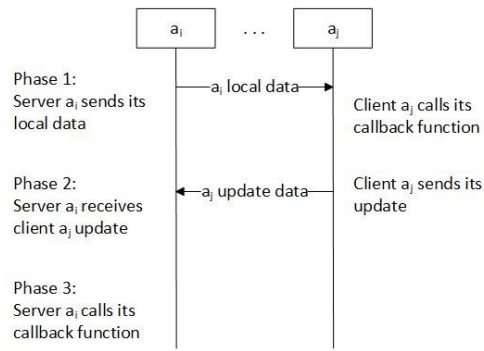

Figure 3: The server-client interaction within a generic FLA.

In case of the centralized one-shot FLA, one of the instances, say $a_1$, is the server, and all other instances, say $a_2$ to $a_n$ are the clients, imagine Figure 3 transformed into a figure where $a_i$ becomes $a_1$ and $a_j$ is replicated $n$-1 times into $a_2$ to $a_n$. So, a1 broadcasts its local data to all the clients and receives back their updates. Plus, in the centralized one-shot FLA there is one important extension of the common operation part illustrated in Figure 3, which is that each client $a_j$ stores its update to its local data after it sent it to the server.

In case of the decentralized one-shot FLA, messages exchanged among the instances comprise the following three fields: the phase number, the source instance id, and the data (payload). The operation from point of view of each instance $a_i$ proceeds through the three phases shown in Figure 3, as follows. In the phase 1, the instance $a_i$ takes the role of a server and sends its data to all other instances, in the phase 2, it takes the role of a client, and in the phase 3, it again takes the role of a server to aggregate updates from all its clients and to store it into its local data. Note that messages may have variable delays and therefore may be received out of the phase order. Therefore, while in phase 2, $a_i$ only processes messages sent in phase 1 and postpones processing of messages sent in phase 2 (by buffering them) until phase 3.

The main difference between the centralized and decentralized one-shot algorithms is that the instance $a_i$, in phase 2, processes a message from the phase 1 by calling its client callback, then sending the update to the source of that message, and finally by skipping storing it locally (in its local data). This means that $a_i$ will calculate the updates for all its servers by using its current local data in phase 2 and will update it only at the end of the phase 3.

## 3 DEVELOPMENT PARADIGM

In this section we present the PTB-FLA development paradigm. The next two subsections present the general concept and the development phases, respectively.

### 3.1 Concept

The PTB-FLA development paradigm is primarily intended to serve as a FLA developer guide through the process of developing a target FLA using PTB-FLA, which we call the FLA development process. The input to this process is the Python sequential program code of target AI/ML algorithm, whereas the output from this process is the PTB-FLA code with the same semantics, which means that for given input data it produces the same output data with some tolerance $e$. The tolerance $e$ is typically some small error value (ideally zero).

Of course, the output PTB-FLA code must be compliant with the PTB-FLA programming model which is a restricted programming model that imposes the following two restrictions: (1) using the Single Program Multiple Data (SPMD) pattern, and (2) specifying code for server and client roles in form of callback functions. Obviously, there are many ways to define such a development process. Our intention was to prescribe it as a paradigm which is much more disciplined than ad hoc development, but also not too rigid to keep it attractive and creative.

The main idea of the PTB-FLA development paradigm was to follow the principle of correct-by-construction, which when applied in this context meant to define the development process that would for a given referent code yield the output PTB-FLA code with the equivalent semantics. Following the approach used by program compilers, we devised the PTB-FLA development paradigm as the series of program code transformation phases where each transformation phase consumes its input code and produces the semantically equivalent code that is closer to the target PTB-FLA code.

### 3.2 Development Phases

There are altogether four development phases, called phase 1, phase 2, phase 3, and phase 4, which are by the convention named by their output code i.e., (1) the referent sequential code, (2) the federated sequential code, (3) the federated sequential code with callbacks, and (4) the PTB-FLA code, respectively.

The input to phase 1 is the row sequential code and the output is the referent sequential code. The input row sequential code may come from various sources and may have various forms. Nowadays, many AI/ML algorithm solutions in Python are available online in Colab notebooks, where snippets of textual mathematical explanations, code snippets, and graphs plots dynamically created by code play (i.e., execution) are interleaved. Typically, these solutions are primarily intended for learning/understanding the solutions through interactive experimentation, where readers are even encouraged to tweak the code and play with it.

To make the output referent sequential code, a PTB-FLA developer essentially must select only the necessary code snippets (leaving out the alternative or redundant snippets), to tweak them if needed, and to integrate them into a standalone Python module(s) that they could preferably run on their PC (localhost), typically in a terminal of some IDE. The important requirement for the referent sequential code is that for a given input dataset it must deterministically produce some output data e.g., learned (trained) model coefficients and/or some quality indicators like accuracy, because this output data is used as the referent data by the next development phases, which they must produce (with some small error) to be semantically equivalent. To that end, a PTB-FLA developer should use asserts that automatically compare whether the output data is (approximately) equal to the referent data, and if not, report the corresponding assertion error.

In phase 2, a PTB-FLA developer makes the federated sequential code by the following three steps: (1) partition the input dataset into partitions (that could be distributed across clients), (2) split the monolithic computing of the complete input dataset into a series of computing on individual partitions (that could be collocated with corresponding partitions) such that they produce the set of partition models, and (3) add the computing for aggregating the set of partition models into the final model and for computing quality indicators (that could be located on a server), as well as for comparing output and referent data. For example, a single function call (calling the function *f*) to process the complete dataset could be split into a series of function calls (calling the same function *f* with different arguments) to process individual dataset partitions. Note that this is still one sequential program that runs on a single machine (developer's PC).

In phase 3, a PTB-FLA developer makes the federated sequential code with callbacks by the following four steps: (1) copy (and tweak) the computing on an individual partition (say a partition number *i*) into the client callback function, (2) replace the series of computing on individual partitions with the series of client callback function calls (with the arguments corresponding to the partition *j* in the call number *j*), (3) copy (and tweak) the computing for aggregating the set of partition models to the final model into the server callback function, and (4) replace the former computing with the server callback function call (the code for computing quality indicators should remain in its place). In the running example, the series of function calls (calling the same function *f* with different arguments) to process individual dataset partitions should be replaced with the corresponding series of client callback function calls, which lead to indirect calls to the function *f* (each call to the client callback function maps to the corresponding call of the function *f*).

In phase 4, a PTB developer makes the PTB-FLA code by the following two steps: (1) add the code for creating the instance *ptb* of the class *PtbFla* and for preparing local and private data for all the instances, and (2) replace the code for calling callback functions (both the series of client callback function calls and the server callback function call) with the call to the function fl_centralized (in case of a centralized FLA) or fl_decentralized (in case of a decentralized FLA) on the object *ptb*.

Generally, a PTB-FLA developer should first develop the centralized FLA and then develop the decentralized FLA, because the centralized FLA is simpler and easier to comprehend. As the next case study shows, when developing the decentralized FLA after the centralized one, a PTB-FLA developer may reuse the code from the first two phases, and then just tweak the callback functions for the last two phases if needed – for example, to get the same output data (i.e., results) by both centralized and decentralized FLAs.

## 4  CASE STUDY: LOGISTIC REGRESSION

In this section we validate and illustrate the PTB-FLA development paradigm by the case study on logistic regression. The input code for phase 1 was the Colab notebook by Adarsh Menon [27], which uses the Social Network Ads (SNA) dataset. SNA dataset comprises 400 samples (or rows) corresponding to user profiles and each record comprises the following features (or columns): (1) User ID, (2) Gender, (3) Age, (4) Estimated (yearly income), and (5) Purchased. Note that in the case study only features Age and Purchased are used.

The main steps in the input code for phase 1 are: (1) split SNA dataset into training and test datasets (320 and 80 samples, respectively), (2) train the linear logistic regression model comprising two coefficients, namely *b0* and *b1*, (3) using the trained model, make predictions whether users in the test dataset would make purchase or not (i.e., whether the predicted probability *p* per user is above the threshold 0.5 or not), and (4) calculate the prediction *accuracy* as the ratio of test users for whom the predictions were correct.

In the next two subsections we apply the PTB-FLA development paradigm to first develop the Centralized Logistic Regression FLA (CLR-FLA, see section 4.1) and then to develop the Decentralized Logistic Regression FLA (DLR-FLA, see section 4.2). Note that algorithms are given in a Pythonic pseudocode.

**4.1 Centralized Logistic Regression FLA**

In the next four subsections we describe the four phases of the PTB-FLA development process that we conducted to develop the centralized logistic regression FLA (phases 1, 2, 3, and 4 are presented in sections 4.1.1, 4.1.2, 4.1.3, and 4.1.4, respectively).

*4.1.1 Phase 1.*

As already mentioned, the input code for phase 1 is taken from [27]. The output code for phase 1 (called the referent sequential code) comprises the main function seq_base_case and two supplementary functions: logistic regression and evaluate, see Algorithm 1 (note: the supplementary functions normalize and predict were not changed and therefore are not included to save space).

ALGORITHM 1: CLR-FLA phase 1 output code

```
// pd is representing the Pandas library
01: seq_base_case()
02:    data = pd.read_csv("Social_Network_Ads.csv")
03:    X_train, X_test, y_train, y_test = train_test_split(data['Age'],
         data['Purchased'], test_size=0.20, random_state=42)
04:    b0, b1 = logistic_regression(X_train, y_train)
05:    y_pred, accuracy = evaluate(X_test, y_test, b0, b1)
// The supplementary function logistic_regression
06: logistic_regression(X, Y, b0=0., b1=0., L=0.001, epochs=300)
07:    X = normalize(X)
08:    for epoch in range(epochs)
09:       y_pred = predict(X, b0, b1)
10:       D_b0 = -2 * sum((Y - y_pred) * y_pred * (1 - y_pred))
11:       D_b1 = -2 * sum(X * (Y - y_pred) * y_pred * (1 - y_pred))
12:       b0 = b0 - L * D_b0
13:       b1 = b1 - L * D_b1
14:    return b0, b1
// The supplementary function evaluate
15: evaluate(X_test, y_test, b0, b1)
16:    X_test_norm = normalize(X_test)
17:    y_pred = predict(X_test_norm, b0, b1)
18:    y_pred = [1 if p >= 0.5 else 0 for p in y_pred]
19:    accuracy = 0.
20:    for i in range(len(y_pred)):
```

21:     if *y_pred*[*i*] == *y_test*.iloc[*i*]:
22:         *accuracy* += 1.
23:     *accuracy* = *accuracy* / len(*y_pred*)
24:     return *y_pred*, *accuracy*

The main function seq_base_case (lines 1-5) takes the following four steps. Step 1 (line 2): load the dataset SNA into the variable *data* of the type Pandas DataFrame. Step 2 (line 3): split the dataset from the variable *data* into the variables *X_train*, *X_test*, *y_train*, and *y_test* of the type Pandas Series, such that test size is 0.2 (or 20%) of the complete dataset (i.e., 80 test samples and 320 training samples), and random splitting always start from the random state 42 (to provide reproducibility of the splitting result and to enable comparing the output data in the next phases with the referent data). Step 3 (line 4): train the model by calling the function logistic_regression on the training data pair (*X_train*, *y_train*) – the return values are the resulting model coefficients: (*b0*, *b1*). Step 4 (line 5): evaluate the model (*b0*, *b1*) on the test data pair (*X_test*, *y_test*) by calling the function evaluate – the return values are the predictions *y_pred* made on *X_test* and the achieved accuracy.

The function logistic_regression (lines 6-14) has four default arguments: *b0*=0., *b1*=0., *L*=0.001, *epochs*=300. The default arguments (*b0*, *b1*) were introduced to enable the so-called incremental training in case when the initial model is given, say by the server; otherwise, the function starts from the default initial model (0., 0.). The default arguments (*L*, *epochs*) are the learning rate and the number of epochs, respectively. The function takes three steps: (1) normalize *X* i.e., *X_train* (line 7), (2) train the model for given number of epochs (lines 8-13), and (3) return the trained model (line 14). The for loop (lines 8-13) comprises three steps: (1) make predictions (line 9), (2) calculate the gradient (*D_b0*, *D_B1*) (lines 10-11), and (3) update the model coefficients (*b0*, *b1*) (lines 12-13).

The function evaluate (lines 15-24) takes four steps: (1) normalize *X_test* (line 16), (2) make predictions (lines 17-18), (3) calculate accuracy (lines 19-23), and (4) return the predications and the accuracy (line 24).

*4.1.2 Phase 2.*

The output code for phase 2 (called the federated sequential code) comprises the main function seq_horizontal_federated (see Algorithm 2) and the supplementary functions from phase 1, and it targets the system with three instances (two clients and one server). We constructed the federated sequential code by following the three general steps for phase 2 in section 3.2, which when applied to the case at hand became: (1) split the training data horizontally (i.e., sample/row-wise) into two partitions with 160 samples each, (2) split a single function call to the function logistic regression into two function per-client function calls, and (3) add the server code for aggregating the two client trained models.

ALGORITHM 2: CLR-FLA phase 2 output code

01: seq_horizontal_federated()
02:     *data* = *pd*.read_csv("Social_Network_Ads.csv")
03:     *X_train*, *X_test*, *y_train*, *y_test* = train_test_split(*data*['Age'],
        *data*['Purchased'], *test_size*=0.20, *random_state*=42)
04:     *X_train_0* = *X_train*.iloc[:160]
05:     *X_train_1* = X_*train*.iloc[160:]

06:    *y_train_0* = *y_train*[:160]
07:    *y_train_1* = *y_train*[160:]
08:    *b00*, *b01* = logistic_regression(*X_train_0*, *y_train_0*)
09:    *b10*, *b11* = logistic_regression(*X_train_1*, *y_train_1*)
10:    *b0* = (*b00* + *b10*)/2.
11:    *b1* = (*b01* + *b11*)/2.
12:    *y_pred*, *accuracy* = evaluate(*X_test*, *y_test*, *b0*, *b1*)

The function seq_horizontal_federated takes six steps. The first two are the same as in the function seq_base_case (lines 2-3). Next steps follow. Step 3 (lines 4-7): split training data into two partitions, more precisely, split *X_train* into *X_train_0* and *X_train_1* (lines 4-5) and *y_train* into *y_train_0* and *y_train_1* (lines 6-7) where suffixes 0 and 1 are indices of client 1 and 2, respectively. Step 4 (lines 8-9): train client models by calling the function logistic_regression on the corresponding training data partitions i.e., (*X_train_0*, *y_train_0*) and (*X_train_1*, *y_train_1*), respectively – the return values are the corresponding model coefficients i.e., (*b00*, *b01*) and (*b10*, *b11*), respectively, where the first index is the model coefficient index and the second index is the client index. Step 5 (lines 10-11): calculate the aggregated model coefficients (*b0*, *b1*). Step 6 (line 12): evaluate the aggregated model by calling the function evaluate on the test data (*X_test*, *y_test*) and the aggregated model coefficients (*b0*, *b1*) – the return values are the predictions *y_pred* made on *X_test* and the achieved accuracy.

For both phase 1 and 2, the achieved accuracy is the same and is equal to 0.9, but the values for the coefficients (*b0*, *b1*) are not equal. Here we use the relative error (absolute value of the difference divided by the true value) as the metric for the quality of the phase 2 output data. The relative error for the phase 2 model coefficients *b0* and *b1* is 8.89%, and 3.75%, respectively. Since the output model accuracy is the same (0.9), these relative errors are acceptable, and therefore we adopted the phase 2 output data (the model coefficients and the model accuracy) as the new referent data for the next two phases i.e., phase 3 and phase 4.

*4.1.3 Phase 3.*

The output code for phase 3 (called the federated sequential code with callbacks) comprises the main function seq_horizontal_federated_with_callbacks (see Algorithm 3) and the supplementary functions from phase 1. We constructed the federated sequential code with callbacks by following the four general steps for phase 3 in section 3.2, which when applied to the case at hand became: (1) copy one of the logistic_regression function calls that operate on an individual training data partitions into the client callback function, (2) replace the series of logistic_regression function calls with the corresponding series of client callback function calls, (3) copy the computing for aggregating the set of partition models to the final model into the server callback function, and (4) replace the former computing with the server callback function call (the evaluate function call should remain in its place).

ALGORITHM 3: CLR-FLA phase 3 output code

01: seq_horizontal_federated_with_callbacks()
02:    *data* = *pd*.read_csv("Social_Network_Ads.csv")
03:    *X_train*, *X_test*, *y_train*, *y_test* = train_test_split(*data*['Age'],

```
            data['Purchased'], test_size=0.20, random_state=42)
04:    X_train_0 = X_train.iloc[:160]
05:    X_train_1 = X_train.iloc[160:]
06:    y_train_0 = y_train[:160]
07:    y_train_1 = y_train[160:]
08:    localData = [0., 0.]
09:    msgsrv = [0., 0.]
10:    msg0 = fl_cent_client_processing(localData,
          [X_train_0, y_train_0], msgsrv)
11:    msg1 = fl_cent_client_processing(localData,
          [X_train_1, y_train_1], msgsrv)
12:    msgs = [msg0, msg1]
13:    avg_model = fl_cent_server_processing(None, msgs)
14:    b0 = avg_model[0]
15:    b1 = avg_model[1]
16:    y_pred, accuracy = evaluate(X_test, y_test, b0, b1)
17:    return [msg0, msg1, [b0, b1]]
18: fl_cent_client_processing(localData, privateData, msg)
19:    X_train = privateData[0]
20:    y_train = privateData[1]
21:    b0 = msg[0]
22:    b1 = msg[1]
23:    b0, b1 = logistic_regression(X_train, y_train, b0, b1)
24:    return [b0, b1]
25: fl_cent_server_processing(privateData, msgs)
26:    b0 = 0.; b1 = 0.
27:    for lst in msgs:
28:        b0 = b0 + lst[0]
29:        b1 = b1 + lst[1]
30:    b0 = b0 / len(msgs)
31:    b1 = b1 / len(msgs)
32:    return [b0, b1]
```

The function seq_horizontal_federated_with_callbacks (lines 1-17) takes 8 steps. The first three are the same as in the function seq_horizontal_federated (lines 2-7). Next steps follow. Step 4 (lines 8-9): prepare the arguments *localData* and *msgsrv* for the following client callback function calls – the former is the client initial model whereas the latter is the message from the server carrying the server initial model. Step 5 (lines 10-11): make the series of two client callback function calls (which replaced the original logistic_regression function calls in lines 8-9 in the phase 2 function seq_horizontal_federated). Note that training data partition is passed through the client callback function argument *privateData* (see line 18). The return values *msg0* and *msg1* are the messages from the clients to the server that carry client updated models that were trained on the client private data. Step 6 (line 12): prepare the argument *msgs* for the following server callback function call – this is the list of messages received from clients carrying their respective updated models (or more briefly called updates). Step 7 (line 13): the server callback function call (which replaced lines 10-11 in the phase 2 function seq_horizontal_federated) – return value *avg_model* is the aggregated model. Step 8 (lines 14-17): unpack the model coefficients *b0* and *b1*, call the function evaluate, and return the list of all the models (this list is useful for the assertion in phase 4).

The function fl_cent_client_processing (lines 18-24) takes 3 steps. Step 1 (lines 19-22): unpack the arguments *privateData* and *msg* into the local variables *X_train*, *y_train*, *b0*, and *b1*, which are needed for the following logistic_regression function call. Step 2 (line 23): make the logistic_regression function call (which is a copy- tweak of the line 8 in the phase 2 function seq_horizontal_federated). Step 3 (line 24): return the client updated model trained on its private data i.e., the client update.

The function fl_cent_server_processing (lines 25-32) takes 2 steps. Step 1 (lines 25-31): aggregate the client models from the list *msgs* (carrying the models from the clients) – the result is the server aggregated model coefficients *b0* and *b1*. Step 2 (line 32): return the final server aggregated model as the list [*b0*, *b1*].

### 4.1.4 Phase 4.

The output code for phase 4 (called the PTB-FLA code) comprises the main function ptb_fla_code_centralized (see Algorithm 4), the supplementary functions from phase 1, and the main and the callback functions from phase 3. We constructed the PTB-FLA code by following the two general steps for phase 4 in section 3.2, which when applied to the case at hand became: (1) add the code for creating the instance *ptb* of the class *PtbFla* and for preparing local and private data for all the instances, and (2) replace the code for calling callback functions with the call to the function fl_centralized on the object *ptb*.

ALGORITHM 4: CLR-FLA phase 4 output code

```
01: ptb_fla_code_centralized(noNodes, nodeId, flSrvId)
02:     data = pd.read_csv("Social_Network_Ads.csv")
03:     X_train, X_test, y_train, y_test = train_test_split(data['Age'],
            data['Purchased'], test_size=0.20, random_state=42)
04:     X_train_0 = X_train.iloc[:160]
05:     X_train_1 = X_train.iloc[160:]
06:     y_train_0 = y_train[:160]
07:     y_train_1 = y_train[160:]
08:     ptb = PtbFla(noNodes, nodeId, flSrvId)
09:     lData = [0., 0.]
10:     if nodeId == 0
11:         pData = [X_train_0, y_train_0]
12:     else if nodeId == 1
13:         pData = [X_train_1, y_train_1]
14:     else
15:         pData = None
```

16:    *ret* = *ptb*.fl_centralized(fl_cent_server_processing,
      fl_cent_client_processing, *lData*, *pData*, 1)
17:    *b0* = *ret*[0]; *b1* = *ret*[1]
18:    *y_pred, accuracy* = evaluate(*X_test, y_test, b0, b1*)
19:    if *nodeId* == *flSrvId*:
20:       *refbs* = seq_horizontal_federated_with_callbacks()
21:       assert *refbs*[2][0] == *b0* and *refbs*[2][1] == *b1*,
      "*b0* and *b1* must be equal to *ref_b0* and *ref_b1*!"
22:    del *ptb*

---

The function ptb_fla_code_centralized (lines 1-22) takes 8 steps. The first three are the same as in the function seq_horizontal_federated_with_callbacks (lines 2-7). Next steps follow. Step 4 (8-15): create the object *ptb* (line 8) i.e., start-up the system, and prepare the local data (line 9) and the private data (lines 10-15) for all the instances. Step 5 (line 16): call the API function fl_centralized on the object *ptb* – the arguments are the callback functions, the local and private data, and the number of iterations (here set to 1 i.e., one-shot execution), whereas the return value is the updated model (client model for client instances and aggregated model for the server instance). Step 6 (lines 17-18): unpack the model coefficients *b0* and *b1*, and call the function evaluate – the return values are the predictions *y_pred* and the achieved *accuracy*. Step 7 (lines 19-21): if the instance is the server, call the function main phase 3 function seq_horizontal_federated_with_callbacks to get the referent output values, and assert that they are equal with the output values produced by this PTB-FLA code. Step 8 (line 22): destroy the object *ptb* i.e., shutdown the system.

This concludes the development of the centralized logistic regression FLA.

### 4.2 Decentralized Logistic Regression FLA

Once we developed a centralized FLA for the system with *n* instances (where *n* > 2), developing its decentralized counterpart for the system with (*n* – 1) instances (note that the server is excluded because it's not needed anymore), which has the same semantics (i.e., it is producing the same output data), is rather straightforward. Obviously, since in the decentralized system, the server is excluded, the remaining peers need to do some extra work to get the same result that the missing server was producing. What is the missing part?

To see the answer, let's focus on the third phase of the generic decentralized FLA, where each peer receives (*n* – 2) updated models from its clients. When compared with the third phase of the generic centralized FLA, where the server receives (*n* – 1) updated models from its clients, we realize that one updated model is missing, and that is the updated model of the peer (in the role of a server) itself. Therefore, the server callback function first must update its local model by training it on its private data (here we can reuse the centralized client callback function), add it at the end of the received list of client models, and then aggregate all the client models, including its own (here we can reuse the centralized server callback function).

This means that we can simply reuse the first three phase of the development process we conducted for the decentralized logistic regression FLA, and then in the fourth phase we need to write the new server callback function and adapt the main function accordingly (see the next subsection on phase 4).

*4.2.1 Phase 4.*

The output code for phase 4 (i.e., PTB-FLA code) comprises the main function ptb_fla_code_decentralized and the new server callback function fl_decent_server_processing (see Algorithm 5), as well as the supplementary functions from phase 1 (section 4.1.1) and the callback functions from phase 3 (section 4.1.3) of the previous CLR-FLA development.

---
ALGORITHM 5: DLR-FLA phase 4 output code
---

01: ptb_fla_code_decentralized(*noNodes*, *nodeId*)
02:    *data* = *pd*.read_csv("Social_Network_Ads.csv")
03:    *X_train*, *X_test*, *y_train*, *y_test* = train_test_split(*data*['Age'],
     *data*['Purchased'], *test_size*=0.20, *random_state*=42)
04:    *X_train_0* = *X_train*.iloc[:160]
05:    *X_train_1* = *X_train*.iloc[160:]
06:    *y_train_0* = *y_train*[:160]
07:    *y_train_1* = *y_train*[160:]
08:    *ptb* = PtbFla(*noNodes*, *nodeId*)
09:    *lData* = [0., 0.]
10:    if *nodeId* == 0
11:      *pData* = [*X_train_0*, *y_train_0*]
12:    else
13:      *pData* = [*X_train_1*, *y_train_1*]
14:    *ret* = *ptb*.fl_centralized(fl_decent_server_processing,
     fl_cent_client_processing, *lData*, *pData*, 1)
15:    *b0* = *ret*[0]; *b1* = *ret*[1]
16:    *y_pred*, *accuracy* = evaluate(*X_test*, *y_test*, *b0*, *b1*)
17:    *refbs* = seq_horizontal_federated_with_callbacks()
18:    assert *refbs*[2][0] == *b0* and *refbs*[2][1] == *b1*,
     "*b0* and *b1* must be equal to *ref_b0* and *ref_b1*!"
19:    del *ptb*
20: fl_decent_server_processing(*privateData*, *msgs*)
21:    *myData* = fl_cent_client_processing(None, *privateData*, [0., 0.])
22:    *msgs2* = *msgs* + [*myData*]
23:    *myData2* = fl_cent_server_processing(None, *msgs2*)
24:    return *myData2*

---

    The main function ptb_fla_code_decentralized was constructed from the function ptb_fla_code_centralized (section 4.1.4) by the following four changes: (1) the argument *flSrvId* is deleted (see lines 1 and 8), (2) the preparation of private data is reduced to preparation for two instances (lines 10-13), (3) the server callback function is changed to fl_decent_server_processing (line 14), and in contrast to the centralized FLA the assert must be satisfied for both instances (line 18).

The function fl_decent_server_processing takes the following three steps: (1) call the centralized client callback function fl_cent_client_processing – the return value is the local model that was updated by training on the private data (line 22), (2) add the updated local model at the end of the list of all client models (line 23), (3) call the centralized server callback function fl_cent_server_processing – the return value is the aggregated model (line 24), and (4) return the aggregated model (line 25).

## 5 CONCLUSIONS

In this paper, we present the PTB-FLA development paradigm. The paradigm comprises the four phases dubbed by the code they produce: (1) the sequential code, (2) the federated sequential code, (3) the federated sequential code with callbacks, and (4) the PTB-FLA code. The PTB-FLA development paradigm is validated and illustrated in the case study on logistic regression, where both centralized and decentralized algorithms are developed.

The main original contributions of this paper are: (1) the PTB-FLA development paradigm, (2) the case study on logistic regression, and (3) the developed centralized and decentralized logistic regression FLAs.

The main advantages of PTB-FLA development paradigm are: (1) it aids the development of FLAs that are correct by construction and (2) it makes the development of FLAs easier. These advantages are achieved by devising the PTB-FLA development paradigm as a series of four program code transformation phases, where each phase produces code semantically equivalent to its input code, and for each development phase its main steps are clearly described.

The potential limitations of the PTB-FLA development paradigm may depend on developers' subjective development experience: (1) for some developers it may be too restrictive, whereas (2) for some other developers it may be too informal. This is because we tried to create a paradigm that is more disciplined than the pure ad hoc approach but not too rigid to let it be attractive and creative.

The main directions of future work are: (1) use the PTB-FLA development paradigm to develop other more complex ML and AI FLAs, (2) continue improving the PTB-FLA development paradigm based on the feedback from developing more complex FLAs, and (3) research adapting and specifying the PTB-FLA development paradigm for AI tools, such as GPT-4, ChatGPT, and alike.

## ACKNOWLEDGMENTS

EU Funded by the European Union (TaRDIS, 101093006). Views and opinions expressed are however those of the author(s) only and do not necessarily reflect those of the European Union. Neither the European Union nor the granting authority can be held responsible for them.